\documentclass[12pt]{article}
\usepackage{amssymb}
\usepackage{amsmath}
\usepackage{graphicx}

\begin{document}

\noindent {\small CITUSC/01-032\hfill \hfill hep-th/0109132 }

{\vskip1.5cm}

\begin{center}
{\Large \textbf{Nonpertubative Effects of Extreme Localization\\[0pt]
in Noncommutative Geometry}}

\bigskip

{\vskip1.0cm}

\textbf{Itzhak Bars}{\footnote{%
This research was partially supported by the US Department of Energy under
grant number DE-FG03-84ER40168.}}

{\vskip1.0cm}

\textsl{CIT-USC Center for Theoretical Physics \& Dept. of Physics and
Astronomy}

\textsl{University of Southern California,\ Los Angeles, CA 90089-2535, USA}

{\vskip1.5cm}

\textbf{Abstract}
\end{center}

{\vskip1.0cm}

``Extremely'' localized wavefunctions in noncommutative geometry have
disturbances that are localized to distances smaller than $\sqrt{\theta },$
where $\theta $ is the ``area'' parameter that measures noncommutativity. In
particular, distributions such as the sign function or the Dirac delta
function are limiting cases of extremely localized wavefunctions. It is
shown that Moyal star products of extremely localized wavefunctions cannot
be correctly computed perturbatively in powers of $\theta $. Nonperturbative
effects as a function of $\theta $ are explicitly displayed through exact
computations in several examples. In particular, for distributions, star
products end up being functions of $\theta ^{-1}$ and have no expansion in
positive powers of $\theta .$ This result provides a warning for
computations in noncommutative space that often are performed with
perturbative methods. Furthermore, the result may have interesting
applications that could help elucidate the role of noncommutative geometry
in several areas of physics.

\newpage

\section{Star-commutators with distributions}

For simplicity we will limit our discussion in this note to a two
dimensional noncommutative plane (generalizations are immediate). The two
noncommutative coordinates are denoted as $x_{1}=x$ and $x_{2}=p$ as a
reminder of the close relation between noncommutative geometry and quantum
mechanics, but we have in mind various applications of noncommutative
geometry in physics, including the quantum Hall effect, strings in large
background fields, and string field theory. The noncommutativity parameter $%
\theta $ has dimensions of ``area'', i.e. units of $x$ times units of $p,$
and its meaning depends on the specific physical application. As we will
make explicit, localization to distances shorter than $\sqrt{\theta }$
produce nonperturbative effects as a function of the parameter $\theta $ in
computations involving the noncommutative geometry.

Consider functions in the noncommutative plane $\Lambda \left( p,x\right) .$
In particular, consider the sign step-function, $\varepsilon \left( p\right)
=\frac{p}{\left| p\right| },$ which takes the values $\pm 1$ for $p\gtrless
0 $ respectively. Its derivative is the Dirac delta function
\begin{equation}
\frac{\partial }{\partial p}\varepsilon \left( p\right) =2\delta \left(
p\right) .
\end{equation}
As is well known, it can be represented as an integral
\begin{equation}
\varepsilon \left( p\right) =\int_{-\infty }^{\infty }\frac{dq}{\pi i}%
e^{iqp}\left( P\frac{1}{q}\right)
\end{equation}
where $\left( P\frac{1}{q}\right) $ is the principal value.

Now consider the Moyal star-commutator of $\varepsilon \left( p\right) $
with any function $\Lambda \left( p,x\right) $. It is given by
\begin{eqnarray}
&&\varepsilon \left( p\right) \star \Lambda \left( p,x\right) -\Lambda
\left( p,x\right) \star \varepsilon \left( p\right) \\
&=&\left\{ e^{\frac{i\theta }{2}\left( \partial _{x}\partial _{p}^{\prime
}-\partial _{p}\partial _{x}^{\prime }\right) }\left[ \varepsilon \left(
p\right) \Lambda \left( p^{\prime },x^{\prime }\right) -\Lambda \left(
p,x\right) \varepsilon \left( p^{\prime }\right) \right] \right\}
_{p=p^{\prime };x=x^{\prime }}  \label{moyal} \\
&=&\varepsilon \left( p-\frac{i\theta \partial }{2\partial x}\right) \Lambda
\left( p,x\right) -\varepsilon \left( p+\frac{i\theta \partial }{2\partial x}%
\right) \Lambda \left( p,x\right) ,  \label{first} \\
&=&\Lambda \left( p^{\prime },x-\frac{i\theta \partial }{2\partial p}\right)
\varepsilon \left( p\right) -\Lambda \left( p^{\prime },x+\frac{i\theta
\partial }{2\partial p}\right) \varepsilon \left( p\right) ,  \label{second}
\end{eqnarray}
where in the last line one sets $p^{\prime }=p$ after the derivatives are
performed.

If one expands any of the expression in Eqs.(\ref{moyal}-\ref{second}) in a
power series in $\theta $ one finds that the result is zero if $p\neq 0.$
This is intuitively understandable, since for $p\neq 0,$ one is trying to
commute $+1$ or $-1$ with some function, and therefore zero appears as a
reasonable result. However, quantum mechanics (or equivalently,
noncommutative geometry) can be tricky because there is a probability
distribution for the values $\pm 1.$ More precisely, every term in the power
series expansion of (\ref{moyal}-\ref{second}) is proportional to the delta
function $\delta \left( p\right) $ or its derivatives (odd number of
derivatives of $\varepsilon \left( p\right) $); therefore, it seems that, if
there is any support for a non-zero result, it is only at $p=0.$ Away from $%
p=0$ the result of the power expansion is apparently zero.

This result correctly applies when $\Lambda \left( p,x\right) $ involves
simple powers of $x.$ Indeed, it is straightforward to use the form of Eq.(%
\ref{second}) to evaluate the commutator when $\Lambda =x,x^{2},x^{3},$ etc.
In such cases the dependence on $\theta $ is necessarily of the perturbative
form. However, it is shown in this note that for more general functions $%
\Lambda \left( p,x\right) $ the perturbative computation described in the
previous paragraph surprisingly misses nonperturbative effects in $\theta $
which are not zero even when $p\neq 0.$ The result of the commutator turns
out to be a smooth function of $\left( p,x,\theta \right) $ that involves
only the inverse powers of $\theta .$

By using the integral representation, the expression in Eq.(\ref{first}) is
evaluated as follows
\begin{eqnarray}
&&\left[ \varepsilon \left( p\right) ,\Lambda \left( x,p\right) \right]
_{\star } \\
&=&\int_{-\infty }^{\infty }\frac{dq}{\pi i}\left( P\frac{1}{q}\right)
\left( e^{iq\left( p-\frac{i\theta \partial }{2\partial x}\right)
}-e^{iq\left( p+\frac{i\theta \partial }{2\partial x}\right) }\right)
\Lambda \left( x,p\right) , \\
&=&\int_{-\infty }^{\infty }\frac{dq}{\pi i}\left( P\frac{1}{q}\right)
\,e^{iqp}\left( \Lambda \left( x+\frac{\theta q}{2},p\right) -\Lambda \left(
x-\frac{\theta q}{2},p\right) \right) .
\end{eqnarray}
The integral is well defined if $\Lambda \left( x,p\right) $ goes to zero
(or even to a constant) at $x\rightarrow \infty .$

Consider the example $\Lambda \left( x,p\right) =f\left( p\right) \left(
1+x^{2}\right) ^{-1},$ with any function $f\left( p\right) $. Then,
according to (\ref{first})
\begin{eqnarray}
&&\left[ \varepsilon \left( p\right) ,\frac{f\left( p\right) }{1+x^{2}}%
\right] _{\star } \\
&=&\int_{-\infty }^{\infty }\frac{dq}{\pi i}e^{iqp}\left( \frac{f\left(
p\right) \left( P\frac{1}{q}\right) }{1+\left( x+\frac{\theta q}{2}\right)
^{2}}-\frac{f\left( p\right) \left( P\frac{1}{q}\right) }{1+\left( x-\frac{%
\theta q}{2}\right) ^{2}}\right) \\
&=&\int_{-\infty }^{\infty }\frac{dq}{\pi i}e^{iqp}\frac{-2\theta xq\left( P%
\frac{1}{q}\right) f\left( p\right) }{\left( 1+\left( x+\frac{\theta q}{2}%
\right) ^{2}\right) \left( 1+\left( x-\frac{\theta q}{2}\right) ^{2}\right) }
\\
&=&\frac{2i\theta xf\left( p\right) }{\pi }\int_{-\infty }^{\infty }\frac{%
dq\,e^{iqp}}{\left( 1+\left( x+\frac{\theta q}{2}\right) ^{2}\right) \left(
1+\left( x-\frac{\theta q}{2}\right) ^{2}\right) }.
\end{eqnarray}
The integral is evaluated by using complex integration, noting that there
are poles\thinspace in the complex $q$\thinspace plane\thinspace at$\,\frac{2%
}{\theta }\left( x\pm i\right) ,\frac{2}{\theta }\left( -x\pm i\right) .$
Closing the contour in the upper half plane (for $p$ positive or zero), or
in the lower half plane (for $p$ negative or zero), and evaluating the
residues of the poles enclosed in either contour, gives the following result
\begin{eqnarray}
\left[ \varepsilon \left( p\right) ,\frac{f\left( p\right) }{1+x^{2}}\right]
_{\star } &=&if\left( p\right) \,\left[ \frac{e^{\frac{2}{\theta }i\left(
x+i\right) \left| p\right| }}{\left( x+i\right) }+\frac{e^{-\frac{2}{\theta }%
i\left( x-i\right) \left| p\right| }}{\left( x-i\right) }\right] \\
&=&\frac{2if\left( p\right) e^{-\frac{2\left| p\right| }{\theta }}}{1+x^{2}}%
\left[ x\cos \left( \frac{2px}{\theta }\right) +\sin \left( \frac{2\left|
p\right| x}{\theta }\right) \right] .
\end{eqnarray}

The significance of this simple exercise is that the result is
nonperturbative in $\theta .$ First of all, it is not zero even when $p\neq
0.$ Second, it is not a power series with positive powers of $\theta .$ It
is still true that as $\theta \rightarrow 0$ the commutator vanishes, but
this happens exponentially, not linearly. A perturbative computation, in
which the Moyal star product in Eq.(\ref{moyal}) is evaluated through a
series expansion in $\theta ,$ misses this result completely as shown above.

Using the Weyl-Moyal correspondence, one may consider the operator image of $%
\varepsilon \left( p\right) .$ Then the computation presented above
corresponds to computing the commutators of this operator with other
operators that act in a quantum mechanical space. Since $\theta $ can be
regarded as $\hbar $ in quantum mechanics, what we have obtained is a
nonperturbative quantum mechanical effect as a function of $\hbar ,$ which
approaches a classical limit as $\hbar \rightarrow 0$, not at the usual
linear rate, but at an exponential rate$.$ There is no usual semi-classical
limit. Certainly this is surprising since a commutator usually (but of
course, not necessarily as seen here) is a power series in $\hbar $ which
starts with the first power$.$

It should be emphasized that $\varepsilon \left( p\right) $ does not decay
at infinity in the noncommutative plane and it is not an infinitely
differentiable smooth function that belongs to $C^{\infty }$. Rather, $%
\varepsilon \left( p\right) $ is a distribution. So, it does not correspond
to a bounded or to a compact quantum operator. The Dirac delta function $%
\delta \left( p\right) ,$ again a distribution, does vanish at infinity. Its
star commutator can be obtained by differentiating the star commutators of $%
\varepsilon \left( p\right) $ before one sets $p^{\prime }=p.$ By
differentiating the results for the example above one obtains
\begin{equation*}
\left[ \delta \left( p\right) ,\frac{f\left( p\right) }{1+x^{2}}\right]
_{\star }=-\frac{2i}{\theta }f\left( p\right) e^{-\frac{2\left| p\right| }{%
\theta }}\,\sin \left( \frac{2}{\theta }x\left| p\right| \right) .
\end{equation*}
We see that the star commutators of $\delta \left( p\right) $ are also
non-perturbative in $\theta $, and non-vanishing even when $p\neq 0.$ By
contrast, the perturbative expansion would have produced a vanishing result
when $p\neq 0.$

\section{Extremely localized wavefunctions}

To understand better how the nonperturbative effects arise, it is
instructive to analyze smeared distributions. Thus, let us consider the
following $C^{\infty }$ functions
\begin{equation}
\delta _{\varepsilon _{1}}\left( p\right) =\frac{e^{-p^{2}/\varepsilon _{1}}%
}{\sqrt{\pi \varepsilon _{1}}},\quad \delta _{\varepsilon _{2}}\left(
x\right) =\frac{e^{-x^{2}/\varepsilon _{2}}}{\sqrt{\pi \varepsilon _{2}}}.
\end{equation}
As long as $\varepsilon _{1},$ $\varepsilon _{2}$ are positive and finite,
these are well behaved, infinitely differentiable, and are representatives
of bounded operators in a quantum Hilbert space according to the Weyl
correspondence. Their star product can be computed by using an integral
representation of the star product \cite{baker}\cite{curtright}, and the
result is a special case of star products of multidimensional gaussians with
matrix insertions and shifts given in \cite{witmoy},
\begin{equation}
\delta _{\varepsilon _{1}}\left( p\right) \star \delta _{\varepsilon
_{2}}\left( x\right) =\frac{1}{\pi \sqrt{\theta ^{2}+\varepsilon
_{2}\varepsilon _{1}}}\exp \left( -\left[ \frac{x^{2}\varepsilon
_{1}+2ixp\theta +p^{2}\varepsilon _{2}}{\theta ^{2}+\varepsilon
_{2}\varepsilon _{1}}\right] \right) ,
\end{equation}
or
\begin{equation}
\delta _{\varepsilon _{2}}\left( x\right) \star \delta _{\varepsilon
_{1}}\left( p\right) =\frac{1}{\pi \sqrt{\theta ^{2}+\varepsilon
_{2}\varepsilon _{1}}}\exp \left( -\left[ \frac{x^{2}\varepsilon
_{1}-2ixp\theta +p^{2}\varepsilon _{2}}{\theta ^{2}+\varepsilon
_{2}\varepsilon _{1}}\right] \right) ,
\end{equation}
and their commutator is
\begin{equation}
\left[ \delta _{\varepsilon _{1}}\left( p\right) ,\delta _{\varepsilon
_{2}}\left( x\right) \right] _{\star }=\frac{-2i\sin \left( \frac{2xp\theta
}{\theta ^{2}+\varepsilon _{2}\varepsilon _{1}}\right) }{\pi \sqrt{\theta
^{2}+\varepsilon _{2}\varepsilon _{1}}}\exp \left( -\left[ \frac{%
x^{2}\varepsilon _{1}+p^{2}\varepsilon _{2}}{\theta ^{2}+\varepsilon
_{2}\varepsilon _{1}}\right] \right) .
\end{equation}
$\delta _{\varepsilon _{1}}\left( p\right) $ becomes the Dirac delta
function $\delta \left( p\right) $ when $\varepsilon _{1}$ approaches zero.
Similarly one may consider independently an $\varepsilon _{2}$ limit to
reach the Dirac delta function $\delta \left( x\right) $.

A perturbative expansion of the results above around $\theta =0$ are
possible. However, these expressions become invalid (not convergent) as soon
as the product $\varepsilon _{1}\varepsilon _{2}$ is smaller than $\theta
^{2}$. Indeed, when $\varepsilon _{1}=0,$ for any finite $\varepsilon _{2},$
the expressions above contain only $\theta ^{-1}$ and cannot have a
perturbative expansion with positive powers of $\theta .$ This shows that
nonperturbative star products are unavoidable for wavefunctions localized to
noncommutative space regions of distances smaller than $\sqrt{\theta }$. In
particular, distributions such as $\varepsilon \left( p\right) ,$ $\delta
\left( p\right) ,$ etc., necessarily have nonperturbative star products as
given in examples above.

As an aside, note that from the expressions above we learn how to
star-multiply Dirac delta functions for mutually noncommuting variables
(when both $\varepsilon _{1}=\varepsilon _{2}=0$)
\begin{equation}
\delta \left( p\right) \star \delta \left( x\right) =\frac{1}{\pi \left|
\theta \right| }\exp \left( -\frac{2ixp}{\theta }\right) ,\quad \delta
\left( x\right) \star \delta \left( p\right) =\frac{1}{\pi \left| \theta
\right| }\exp \left( \frac{2ixp}{\theta }\right) .
\end{equation}

Similarly, the two dimensional Dirac delta function in noncommutative space
can be obtained from the smeared distribution
\begin{equation}
\delta _{\varepsilon }^{\left( 2\right) }\left( x,p\right) =\delta
_{\varepsilon }\left( x\right) \delta _{\varepsilon }\left( p\right) =\frac{1%
}{\pi \varepsilon }\exp \left( -\frac{x^{2}+p^{2}}{\varepsilon }\right) .
\end{equation}
The star product of two such gaussians with different $\varepsilon
_{1},\varepsilon _{2}$ was given in \cite{bayen}\cite{baker} and again is a
special case of the results given in \cite{witmoy}
\begin{equation}
\delta _{\varepsilon _{1}}^{\left( 2\right) }\left( x,p\right) \star \delta
_{\varepsilon _{2}}^{\left( 2\right) }\left( x,p\right) =\frac{1}{\pi
^{2}\left( \theta ^{2}+\varepsilon _{2}\varepsilon _{1}\right) }\exp \left( -%
\frac{\varepsilon _{2}+\varepsilon _{1}}{\theta ^{2}+\varepsilon
_{2}\varepsilon _{1}}\left( x^{2}+p^{2}\right) \right) .
\end{equation}
As before, if $\varepsilon _{1}\varepsilon _{2}$ is small compared to $%
\theta ^{2},$ nonperturbative effects take over. In particular, for $%
\varepsilon _{1}=0,$ at any $\varepsilon _{2},$ the result is purely a
function of $\theta ^{-1}.$ Their commutator $\left[ \delta _{\varepsilon
_{1}}\left( x,p\right) ,\delta _{\varepsilon _{2}}\left( x,p\right) \right]
_{\star }=0$ is evidently zero for all $\varepsilon _{1},\varepsilon _{2},$ $%
\theta .$ A byproduct of this exercise is the following formula for the star
product of two 2-dimensional delta functions in noncommutative space (for $%
\varepsilon _{1}=\varepsilon _{2}=0$)
\begin{equation}
\delta ^{\left( 2\right) }\left( x,p\right) \star \delta ^{\left( 2\right)
}\left( x,p\right) =\frac{1}{\pi ^{2}\theta ^{2}}
\end{equation}
Multi-dimensional generalizations, and more complicated examples can be
easily computed by using the general star product formulas for generating
functions given in \cite{witmoy}.

\section{Comments}

Various distributions may well play a role in a physical setting that
involves non-commutative geometry, just as they do in commutative geometry.
We have learned in this note that one should expect nonperturbative behavior
in the star products of the distributions $\varepsilon \left( p\right) ,$ $%
\delta \left( p\right) ,$ $\delta \left( x\right) ,$ etc., and of course,
this would extend to their derivatives. The star algebra of distributions
with functions and with other distributions can be computed by using similar
methods, and we expect to find nonperturbative behavior in general in such
star products.

In the noncommutative geometry that arises in string theory \cite{ncstring},
$\theta $ is proportional to the inverse of a large background antisymmetric
field. In the quantum Hall effect \cite{qhe}, $\theta $ is proportional to
the inverse of the background magnetic field. In string field theory, having
realized that the string star product is basically the Moyal star product
\cite{witmoy}, we see that $\theta $ is determined by the fundamental string
length (since $\Delta p\sim \Delta x/\alpha ^{\prime }$ in string theory).
Also, as already mentioned, $\theta $ is $\hbar $ in quantum mechanics.
Nonperturbative behavior in such parameters would be of great interest, and
we expect it to be relevant when wavefunctions probe distances shorter than $%
\sqrt{\theta }$.

The non-perturbative effect discussed here is intriguing, and one wonders if
it has interesting applications in various areas of physics? If so, it could
help elucidate the content and role of noncommutative geometry in physics.

\section{Acknowledgements}

I thank M.M. Sheikh-Jabbari for a stimulating remark that led to this
investigation and for commenting on the manuscript.

\bigskip


\begin{thebibliography}{9}
\bibitem{baker}  G. Baker, Phys. Rev. \textbf{105} (1958) 2198.

\bibitem{curtright}  C. Zachos, ``A survey of star product geometry'',
hep-th/0008010; T. Curtright, T. Uematsu, C. Zachos, ``Generating all Wigner
Functions'', J. Math. Phys. \textbf{42} (2001) 2396 or hep-th/0011137.

\bibitem{witmoy}  I. Bars, ``Map of Witten's $\star $ to Moyal's $\star ",$
hep-th/0106157, to appear in Phys. Lett.; For the computations relevant to
the present paper one specializes the formulas in Eqs.(48-57) to the
appropriate 2$\times $2 matrices $M_{1},M_{2}$, and take vanishing shifts $%
\lambda _{1}=\lambda _{2}=0,$ while the $\theta $ dependence is inserted by
replacing $\sigma $ in Eq.(57) with $\theta \sigma .$

\bibitem{bayen}  F. Bayen, M. Flato, C Fronsdal, A. Lichnerowicz, D.
Sternheimer, Ann. Phys. \textbf{111 }(1978) 111.

\bibitem{ncstring}  For a review, see M. Douglas and N. Nekrasov,
``Noncommutative field theory'', hep-th/0106048.

\bibitem{qhe}  L. Susskind, ``The Quantum Hall fluid and noncommutative
Chern-Simons theory'', hep-th/0101029.
\end{thebibliography}
\end{document}